\documentclass[epj]{svjour}
\usepackage{graphics}
\begin{document}
\title{Pentaquark Searches}
\subtitle{(A Lesson in Statistics)}
\author{Ken Hicks\inst{1}
% \thanks is optional - remove next line if not needed
\thanks{\emph{Supported in part by NSF grant PHY-0555558.}}
}                     % Do not remove
\institute{$^1$Department of Physics \& Astronomy, Ohio University, Athens OH 45701 USA}
\date{Received: Jan. 17, 2007 / Revised version: March 7, 2007 }
% The correct dates will be entered by Springer
%
\newcommand{\thplus}{$\Theta^+$}
\abstract{
Current evidence does not favor the existence of the \thplus~pentaquark, 
which was reported by several groups in the years after 2002. The 
question naturally arises: how could many groups could have seen 
fluctuations in their data at the level of 3-5 $\sigma$ statistical 
significance? An example of a statistical fluctuation is given and 
the number of $\sigma$'s necessary for claims of discovery are 
examined.  Using this guideline, a possible answer to the above 
question is presented.
 % end of PACS codes
} %end of abstract
\maketitle
\section{Introduction}
\label{intro}

The \thplus~pentaquark, if it exists, has a predicted \cite{diakonov} 
quark structure of $udud\bar{s}$ and a narrow width.  
Recent theoretical work \cite{weigel}, as an alternative to 
Ref. \cite{diakonov}, gives a modern estimate of its width.
The question of whether the \thplus~exists or not is an important 
one for quantum chromodynamics (QCD). 
In particular, if the \thplus~were to exist, then non-perturbative 
QCD would require quark-quark correlations (or some other mechanism) 
that would prevent the \thplus~from quickly falling apart into lighter 
components (a kaon plus a nucleon) \cite{jaffe}. Hence, the simplest 
quark model predicts pentaquarks with a wide width, which would be 
difficult to observe in experiments.

The observation of a potentially narrow pentaquark by the LEPS 
Collaboration \cite{nakano} at a mass of 1540 MeV was initially 
greeted with some skepticism.  
If true, this would suggest that the color-magnetic attraction 
between quarks could be stronger than previously thought\cite{karliner}.  
Other experiments set out to either confirm or deny the existence 
of the \thplus~using data already available. 
Experiments from DIANA \cite{diana}, CLAS \cite{clas-g2,clas-g6} 
and SAPHIR \cite{saphir} found positive evidence for the \thplus, 
seeming to confirm the LEPS result. Other positive results followed 
within the year, as well as null results (for a review, see 
Ref.~\cite{hicks}).  The conflicting positive and null evidence 
were confusing, and clearly new experiment with better statistics 
were needed to provide answers.

Next, the statistical significance of the positive evidence for the 
\thplus is reviewed.  Then data with higher statistics will be examined, 
and the possibility of repeating some previous experiments with 
more statistics will be discussed.

\section{Statistical Significance}

Soon after evidence for the \thplus~was published, inconsistencies in 
the mass measured by different experimental groups were noticed.  In 
addition, the statistical significance of the \thplus~peaks was 
questioned, due to uncertainties in the shape of the background 
under the ``peaks".  A nice review of these uncertainties was 
presented by Pochodzalla \cite{josef}, where the data was plotted 
with error bars and without fitted curves.  The reader is encouraged 
to read Ref. \cite{josef} for more details.

\begin{figure}
\resizebox{0.5\textwidth}{!}{
  \includegraphics{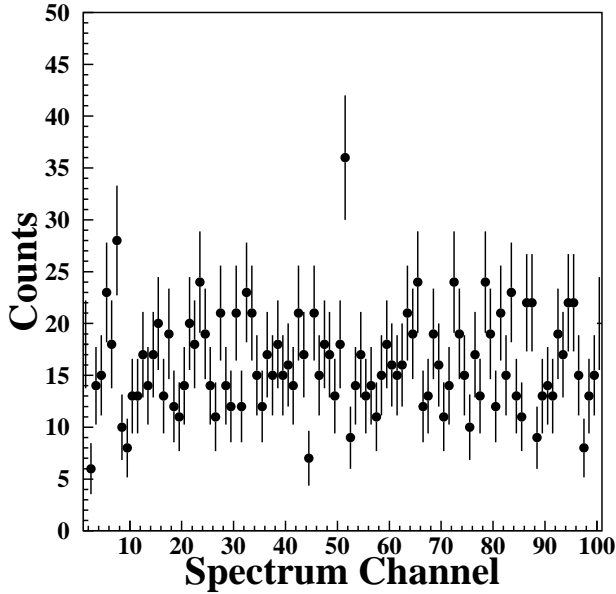}
}
\caption{Example of a Monte Carlo spectrum with a flat distribution 
of the spectrum and a mean value of 15 counts.}
\label{fig:1}
\end{figure}

Here, I will take a slightly different approach.  My goal is to 
determine what level of statistical significance is sufficient before 
one can reasonably claim that a new particle has been seen.  The 
simplest case to study is a Monte Carlo spectrum generated with a 
flat background, as shown in Fig. \ref{fig:1}.  Your eye is drawn 
to the point in the middle, which is unusually high.  What is the 
statistical significance of this ``peak"?  This question is not 
well formed, so let us examine two ways to ask a better question.

First, we will assume that the ``peak" in Fig. \ref{fig:1} is a 
fluctuation of the uniform background.  A fit to the background 
gives a value of $B=15.0 \pm 0.4$ counts.  The signal above 
background is $S=36-15=21$ counts.  The significance of the background 
fluctuation is \cite{frodesen}, in units of standard deviations:
$$ S/\sqrt{B+V} $$
where $V$ is the variance in the background.  In this case, the 
variance $V$ is small, and the significance is, to good approximation, 
$21/4 \simeq 5 \sigma$. In other words, the background has a standard 
deviation, $\sigma_B \simeq \sqrt{B}$ of $\pm 4$ counts, and the 
fluctuation of the ``peak" is five times this standard deviation.  
The probability of a $5\sigma$ fluctuation is about $5\times 10^{-5}$, 
or about once per half-million cases.  
Indeed, this is easy to verify by Monte 
Carlo, by generating a million spectra and seeing how many times 
the ``peak" exceeds 20 counts above background.  Of course, the 
fluctuation can happen anywhere in the spectrum, which has 100 
channels, so the probability of a $5\sigma$ fluctuation {\it anywhere 
in the spectrum} is only once per 200 cases.

The above discussion is rather pedantic, but for a reason. 
There are some publications in the \thplus~literature 
where the statistical significance is not properly calculated. 
The primary reason is the denominator in the above equation, $B+V$. 
In the Monte Carlo example, the value of 
$B+V$ was clear, but in real data the shape of the background is 
often uncertain. So the variance can be large.  

Second, let us assume that the ``peak" is a real signal.  In this 
case, the Monte Carlo generator must be modified, so that the 
background is generated independently from the peak.  In this 
case, the statistical significance is determined by the uncertainties 
in both the signal and the background. The uncertainty in the 
background is $\sigma_B \simeq 4$ counts.  The uncertainty in the 
signal is $\sigma_S = \sqrt{21} = 4.5$ counts.  Adding these in 
quadrature gives $\sigma_{S+B} = 6.0$ counts. Now the statistical 
significance is $21/6 = 3.5\sigma$. This roughly agrees 
with an ``eyeball" estimate from Fig. \ref{fig:1}, where the 
error bar on the ``peak" is about $\pm 6$ counts, and the backgound 
is lower by the length of about 3 times the error bar. However, 
the probability of a $3.5\sigma$ fluctuation is about one 
chance out of 1000. In other words, the probability of a real 
signal to fluctuate {\it down} into the level of the background 
is different than the probability of the background to 
fluctuate {\it up}, as shown in Fig. \ref{fig:1}.

The main point of this discussion is that your eyes can be deceiving 
when estimating statistical significance, and one must rely on 
solid statistical analysis to determine the significance of a signal. 
Perhaps the best way to calculate the significance is: 
(1) fit the shape of the background without the region of the 
``peak", (2) use a Monte Carlo to generate random fluctuations 
around the background shape, (3) determine the number of times 
an excess is found in the ``peak" region, (4) calculate the 
probability of a {\it fluctuation of the given background}, (5) 
repeat this for several reasonable variations of the background.
By applying this procedure to some of the \thplus~data, it is 
easy to see that a $5\sigma$ signal can easily become a $3\sigma$ 
signal for a different background shape 
(for example, see Ref. \cite{mckinnon}).

Of course, there are other complications in addition to those 
mentioned above.  Arbitrary cuts in the data analysis should be 
avoided at all costs.  Each analysis cut should be justified 
independent of its effect on the final mass spectrum. Pessimistic 
estimates of the background should be shown in addition to the 
best-fit background shape.  The data should also be examined with 
different binning, or alternatively use the log-likelihood analysis 
that is described in many textbooks.  These techniques were 
used in some, but not all, of the \thplus~publications.  The 
bottom line is that it takes time and effort to do a good 
statistical analysis, and even then it may result in an estimate 
of the statistical significance of a peak that is too optimistic.

So how many $\sigma$'s are needed to claim discovery?  This question 
has been discussed by Ref. \cite{frodesen}, where they argue that 
a $5\sigma$ peak is not so uncommon, especially if 
there are several cuts (or event selections) applied to the data. 
In addition, an experimentalist may look at hundreds of spectra 
in the course of an analysis, and each spectrum may have hundreds 
of channels.  As a result, I suggest that a statistical significance 
of $7\sigma$ is a reasonable baseline for claims of discovery. 
Even so, a single instance of $7\sigma$ is not convincing, and 
confirmation by an independent experiment {\it at the same level} 
of statistical significance is desired.  This latter point is 
emphasized in the \thplus~review by the Particle Data Group 
\cite{trilling}.

In summary, a few results of only 4-5$\sigma$ are not enough to claim 
discovery.  Few people would be convinced that the ``peak" in 
Fig. 1 is real, unless it could be reproduced with higher statistics. 

\section{ Review of \thplus~Experiments }

The first five publications showing positive evidence for the \thplus~are 
given in column 1 of Table \ref{tab:1}.  The collaborations 
are: LEPS \cite{nakano}, Diana \cite{diana}, CLAS \cite{clas-g2}, 
SAPHIR \cite{saphir} and HERMES \cite{hermes}.  The quoted statistical 
significance in these papers is about 4-5 $\sigma$, although the 
uncertainties in the background under the peaks suggests that 
the statistical significance should have been smaller.  Each of 
these experiments has been repeated, sometimes by other groups, 
although the experimental conditions were not reproduced exactly. 
The ``repeat" experiments are listed in column 4 of Table 
\ref{tab:1}.

The LEPS collaboration repeated their earlier experiment, except 
using a deuterium target rather than Carbon.  Because the liquid 
deuterium target is longer than the compact carbon target, one of 
the analysis cuts from the original analysis \cite{nakano} (using 
vertex tracking to reduce background) was no longer possible. However, 
the statistics in the $K^-$ missing mass spectrum was more than 5 
times higher.  The resulting spectrum has been published only in 
a conference proceeding \cite{meson04} and shows a peak at the 
same mass as before.  The statistical significance of this signal 
was not reported, but is likely in the range of 3-5 $\sigma$. The 
LEPS group decided to take more data on deuterium before publishing 
the results. 

It is perhaps worth noting that the LEPS result is the only one 
of the original five publications where the \thplus~peak has been 
seen in the repeated measurement.  
Either this means that the \thplus~is very peculiar, requiring a 
special kinematic range to be produced, or some other explanation 
is necessary, such as a kinematic reflection from another reaction.  
Kinematic reflections have been investigated by the 
LEPS group, but there is no obvious way that this could happen.
Other explanations, such as excessive data cuts, do not make sense 
because only basic cuts (such as particle ID and $\phi$ exclusion) 
are applied, and variations of the cut limits do not erase the peak.  
However, if the \thplus~is not seen elsewhere then the LEPS data alone 
are not sufficient to claim discovery.

The DIANA results \cite{diana} are still viable, but just barely. 
These data are from old bubble chamber experiments using a $K^+$ 
beam on Xenon. Cuts are applied to reduce the background from 
kaon charge exchange.  Their results were not reproduced directly, 
but are severely limited by analysis from the Belle Collaboration. 
For the Belle results, a kaon was tagged from D-meson decay, which 
interacts with Silicon in their vertex detector, followed by 
detection of $pK^0$ and $pK^+$ pairs.  They estimate and subtract 
events from kaon charge exchange.  The resulting mass spectrum 
does not show any \thplus~peak.  Belle's upper limit for \thplus~
production is below that calculated from the DIANA experiment, 
but still within one standard deviation of the DIANA result.

While the Belle result does not entirely rule out the DIANA
result, it puts a severe limit on the possible width of the 
\thplus~at less than 1 MeV.  Such a narrow width is difficult 
to reconcile with theoretical estimates \cite{ellis}, but still 
on the edge of possibility \cite{weigel,carlson}.  At the very least, 
the DIANA result should be regarded with great caution.

The original CLAS result used older data (called g2a) from deuterium 
that was analyzed quickly after the LEPS result was announced. The 
reaction is exclusive, where all momentum and energy is accounted 
for in the final state. However, this could only be done if the 
proton, which would normally be a spectator, was kicked out of 
the target due to a two-step process. By studying the mirror 
reaction that produces the $\Lambda$(1520), via the 
$\gamma d \to K^+ \Lambda^* n$ reaction, two-step reactions 
accounted for about half the photoproduction cross section there. 
The two-step mechanism, required to detect the proton in the 
\thplus~reaction has both good and bad aspects. The good side is 
that an exclusive reaction can be measured.  The bad side is that 
the amplitude for this reaction is difficult to calculate.

\begin{figure}
\resizebox{0.5\textwidth}{!}{
  \includegraphics{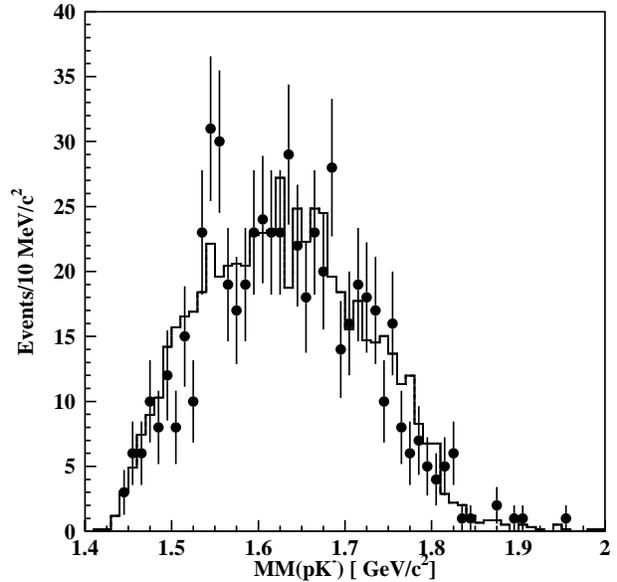}
}
\caption{Comparison of the original CLAS data \cite{clas-g2} 
(points with error bars) and the repeat measurement \cite{mckinnon}
(solid histogram, scaled down by a factor of 5.92). }
\label{fig:2}
\end{figure}

The repeat of the CLAS result is shown in Fig. \ref{fig:2} by 
the solid histogram (which has been rescaled down by a factor 
of 5.92) overlayed on top of the original ``g2a" data. In this 
comparison, the photon energy range was constrained to be the 
same in both analyses.  The new result shows that the shape 
of the background differs from the one expected from the 
g2a data.  In particular, the g2a points on either side of the 
``peak" drop below the histogram. Using the new data, the 
g2a ``peak" is about a $3\sigma$ fluctuation \cite{mckinnon}.  
However, the complication of the two-step reaction prevents 
the new data from placing strong limits on the fundamental 
(one-step) \thplus~cross section.  An upper limit of 3 nb on 
the total cross section was found \cite{mckinnon}, but this 
number depends on the rescattering model used.

The SAPHIR Collaboration were the first to publish a \thplus~search 
using the $\gamma p \to K^+K^0n$ reaction \cite{saphir}.  Using a 
proton target essentially eliminates two-step reactions and hence 
this measurement would be more convincing if the \thplus~were seen.
The SAPHIR experiment used a kinematic fit to determine the 
particle identification (PID).  
Because their detector had a short path for time-of-flight, 
detection of kaons was difficult.  
However, they saw a clear peak for the $\Lambda$(1520), 
showing that their PID was good enough for this purpose.  
After applying a forward-angle cut on the $K^0$'s, they found 
a \thplus~peak with a cross section that was a substantial 
fraction of the $\Lambda$(1520) cross section.

One year later, the CLAS Collaboration measured the same reaction as 
SAPHIR, but with more than 10 times higher statistics \cite{clas-g11}.  
The resulting \thplus~mass spectrum was completely void of peaks, 
even when SAPHIR's data cuts and photon energy range were applied. 
An upper limit of 1 nb was found for the total cross section, which 
is over 100 times smaller than the $\Lambda$(1520) cross section.
Clearly, the SAPHIR result could not be reproduced.

The HERMES Collaboration \cite{hermes} used high-energy $e^+$ on a fixed 
deuteron target to produce events with a $K^0$ and a proton. The 
invariant mass spectrum for $K^0p$ pairs showed a peak near 1528 MeV, 
about 10 MeV lower than the \thplus~peak from other experiments. 
Their analysis was admirable in working to quantify the 
background under their peak, and a good job in estimating the 
statistical significance.  

The BaBar experiment, using colliding $e^+e^-$ beams, had some 
beam halo hit the Berilium beam pipe in their detector.  The resulting 
luminosity was large, and hence they had data for high-energy $e^+$ 
on a fixed Be target with high statistics.  Their results have not 
yet been published, but were reported by Dunwoodie \cite{dunwoodie} 
where preliminary results showed a smooth spectrum without any 
\thplus~peak.  Although there are questions in comparing the BaBar 
results to those of HERMES, it seems that again the \thplus~
could not be reproduced.  

Having seen that the \thplus~could not be reproduced in four out 
of five attempts to confirm the earlier evidence in independent 
experiments, it only makes sense to be highly suspicious of the 
existence of the \thplus.
Does this mean that experimentalists should stop searching for 
pentaquarks?  My answer is no, for the reasons given in the next 
section.  But clearly mistakes were made in the statistical 
significance estimates of some early searches and 
extra effort is necessary to guard against future over-optimistic 
estimates.

\section{ Pending Searches }

There are several pentaquark searches that have positive results 
with high statistical significance, but have not yet been repeated. 
These experiments are listed in Table \ref{tab:2} for the collaborations 
ITEP \cite{itep}, ZEUS \cite{zeus}, CLAS \cite{clas-g6} and 
COSY/TOF \cite{cosy}. Two of these (ZEUS and COSY) have new data 
with higher statistics, but no results yet.
The one with the highest estimated significance (CLAS) requires 
a higher energy beam than Jefferson Lab can presently extract, 
but will likely be repeated in 2008 (as part of a run that 
will also measure meson production).

The logic of why the pentaquark searches should be continued is 
obvious.  If the pentaquark exists, then this has important 
consequences for non-perturbative QCD.  The \thplus~has not been 
seen in some experiments, but to conclude that the \thplus~will 
not be seen in {\it any} experiment is illogical. Having said that, 
it is also true that it is {\it unlikely} that the \thplus~peaks
seen in the experiments of Table \ref{tab:2} will be reproducible, 
based on the irreproducibility of several experiments in Table 
\ref{tab:1}. However, if the physics is important, then the data 
should not be left in this confused state.  The experiments should 
be repeated and either confirmed or rejected.

Since no new results are yet available for the experiments listed 
in Table \ref{tab:2}, let us take a closer look at the published 
results.  The ITEP group \cite{itep} analyzed neutrino-nucleus 
interactions in bubble chamber data, and formed the invariant mass 
for $K^0p$ pairs.  Looking at their data, I estimate about 17 counts 
in the peak on top of a background of about 15 counts. Using the 
equation above, with $V=0$, gives 4.4$\sigma$, far below the 
estimate of 6.7$\sigma$ claimed in their paper. The significance is 
likely lower, since the variance is certainly bigger than zero.

The ZEUS results \cite{zeus} require a cut on $Q^2 > 20$ GeV to 
see a peak.  It is not clear why this kinematic cut is necessary, 
but from analysis of other known resonances, the ZEUS group claims 
that better resolution is obtained in their detector when this 
cut is used.  However, a bigger question is why other high-energy 
experiments such as CDF, HERA-B or FOCUS (see Ref. \cite{hicks} 
for a summary) do not see any \thplus~peak in their data?  
If the ZEUS result is correct, then this argues that 
the \thplus~can be produced by fragmentation, which would be seen 
in many high-energy experiments.  This suggests that the peak 
seen by ZEUS may not be the \thplus~but some other feature of their 
data.

The CLAS results \cite{clas-g6} in Table \ref{tab:2} are different 
from the two CLAS experiments in Table \ref{tab:1}. The former 
measured the $\pi^+K^-K^+p$ final state, with a cut on $\pi^+$ 
backward of about 37$^\circ$ in the center-of-mass frame. 
Without this cut, the \thplus~peak cannot be seen.
See Ref. \cite{clas-g6} for a justification of this cut.
Looking at their data,
I estimate a statistical significance of 4-5$\sigma$ rather than 
the 7-8$\sigma$ given in Ref. \cite{clas-g6}. This is still 
a significant result, and it will be interesting to see whether 
their peak is reproducible when they take more data next year.

The COSY result is perhaps the most intriguing of the 
results in Table \ref{tab:2}. It is the only one where 
a hadronic proton-proton reaction is used.  This has an advantage 
as there is almost no background, since the reaction is so close to 
threshold. Although some people have questioned their particle 
identification scheme, the confusion is likely due to the fact 
that they use a purely geometric technique, which is different 
from that used by many other experiments. 
A close examination of their data 
analysis has convinced me that it is reasonable.  The key thing 
about the COSY data is that they have a new data set with more 
than five times higher statistics. If they can 
reproduce their \thplus~peak, then this will generate a lot of 
interest (and a lot of questions) in the nuclear physics community.

I have not mentioned all experiments with evidence 
for (or against) the \thplus, but a comprehensive review is 
beyond the scope of this article. Instead, I have chosen to 
mention only experiments that I know have been repeated or are 
likely to be repeated in the near future.

\section{ Theoretical Developments }

With the new negative evidence from the ``repeat" experiments 
in Table \ref{tab:1}, the question arises as to whether the 
\thplus~is still theoretically viable.  In other words, is there 
a theory that is consistent with the negative evidence and also 
the potentially positive results (from LEPS)? The theory calculation 
by Nam, Hosaka and Kim \cite{nam} answers in the affirmative, 
provided that the \thplus~has spin-parity $J^\pi = 3/2^+$.

In Ref. \cite{nam}, they argue that for \thplus~production, the 
$s$- and $u$-channel diagrams are suppressed compared with the 
$t$-channel and the contact term. They also note that the contact 
term is present only for production on the neutron. For $J=3/2$ 
in their model, they find that the contact term dominates by a 
factor of 25-50 over $K^*$ exchange in the $t$-channel.  Hence 
it is possible to see the \thplus~at LEPS (for production from 
the neutron) and not see it for the SAPHIR reaction (using a 
proton target), provided the width of the \thplus~is 1 MeV or less. 
Furthermore, their calculation for $J^\pi = 3/2^+$ gives almost 
all of the cross section at forward kaon angles, where the CLAS 
data \cite{clas-g2} has a hole in the acceptance.

Although the above model is consistent with the negative evidence 
from CLAS (both proton \cite{clas-g11} and deuteron \cite{mckinnon} 
data), it still does not explain all of the negative evidence from 
high-energy experiments.  If the \thplus~exists, then it would 
be necessary for its production to be strongly suppressed in 
fragmentation processes.  Using QCD string theory, one model
\cite{suganuma} shows that it is very difficult to produce the 
\thplus~in fragmentation. 

With advances in lattice gauge theory, one can also ask whether 
lattice calculations see any hint of the \thplus.  The answer 
is yes, but only for spin-parity $J^\pi=3/2^+$.  The evidence 
is presented in Ref. \cite{lasscock} (see Fig. 9 of their paper).
The resonance signature, which they have seen for other known 
baryon resonances, is for the mass to drop below that of the 
scattering states, which is seen only for $J^\pi = 3/2^+$.
The authors also show that there is no resonance signal for 
$J^\pi = 1/2^\pm$ or for $3/2^-$. 

%\begin{figure}
%\resizebox{0.5\textwidth}{!}{
%  \includegraphics{lasscock.ps}
%}
%\caption{ Lattice results for a pentaquark with isospin $I=0$ 
%and positive parity (solid points) compared with S-wave and 
%P-wave meson-baryon scattering states, from Ref. \cite{lasscock}. 
%Only in the case of $J^\pi=3/2^+$ does the mass of the pentaquark 
%drop below the scattering states, which is the signature of a 
%baryon resonance.}
%\label{fig:3}
%\end{figure}

It is worth mentioning that there is one other lattice result 
for $J=3/2$, done by a Japanese group \cite{ishii}.  Their 
result has fairly large effective quark masses, and their 
effective mass plots for positive parity have a strange shape 
(see Fig. 3 of Ref. \cite{ishii}) and 
so it is not clear whether their null results are more valid than 
those of Ref. \cite{lasscock}.

The bottom line is that it is still possible for the \thplus~to 
be consistent with all existing data, provided it has $J^\pi=3/2^+$ 
and a narrow width. This would make the \thplus~a very peculiar 
particle, but not yet outside the realm of possibility. 

\section{ Summary }

Using a Monte Carlo spectrum as an example, we see that a 4$\sigma$ 
peak is not very convincing. Furthermore, several 4-5$\sigma$ 
results should not be taken too seriously.  It was suggested that 
two independent results of 7$\sigma$ or more is the threshold 
where discovery can be claimed.  Experimentalists should aspire 
to making pessimistic, not optimistic, estimates of the statistical 
significance of a mass peak.

Of the first five publications showing positive evidence for the 
\thplus, only one (LEPS) has been able to reproduce a peak in a 
``repeat" measurement.  However, the second LEPS result has been 
presented only at conferences, and has not been formally published.  
The bottom line is that the \thplus~has not stood up to the scrutiny 
of higher-statistics measurements.  Several other \thplus~experiments  
have yet to be repeated.

The outlook for the \thplus~looks bleak, but this does not mean that 
we should abandon pentaquark searches.  At present, the data are 
still contradictory, and experimental facilities should endeavor to 
pursue the outstanding claims and put this question (whether the 
\thplus~exists or not) to rest.  I suggest that we be patient yet 
cautious as the experiments make progress.
%
% Format for books
% Author, \textit{Book title} (Publisher, place year) page numbers
\newcommand{\etal}{ {\it et al.}, }

\onecolumn
\begin{table}
\caption{The first five positive \thplus~publications and the 
results of repeat measurements with higher statistics. The 
column labeled $\sigma$ show the statistical significance quoted 
in the publication. 
The column labeled ``Increase" shows the factor by which the 
number of counts in the mass spectrum increased.}
\label{tab:1}       % Give a unique label
\begin{center}
\begin{tabular}{lcc|lccc}
\hline\noalign{\smallskip}
\multicolumn{3}{c|}{Original Experiment} & 
\multicolumn{4}{c}{Repeat Measurement} \\
\noalign{\smallskip}\hline\noalign{\smallskip}
Group & Reaction & $\sigma$'s & Group & Reaction & Increase & Result \\
\noalign{\smallskip}\hline\noalign{\smallskip}
LEPS & $\gamma C \to K^+K^-X$	& $\sim 4$ & 
LEPS & $\gamma d \to K^+K^-X$	& $\sim 5$ & $\sim$ 3-5$\sigma$ \\
DIANA& $K^+Xe \to K^0 p X$	& $\sim 4$ &
Belle& $K^+Si \to K^0 p X$	& $\sim 10$& $\Gamma_{\Theta^+} < 1$ MeV \\
CLAS & $\gamma d \to K^+K^-pn$	& $\sim 5$ &
CLAS & $\gamma d \to K^+K^-pn$	& $>6$     & $\sigma_{tot} < 3$ nb \\
SAPHIR&$\gamma p \to K^+K^-n$	& $\sim 5$ &
CLAS & $\gamma p \to K^+K^-n$	& $>10$    & $\sigma_{tot} < 1$ nb \\
HERMES&$e^+d \to K^0 p X$	& $\sim 4$ &
BaBar& $e^+Be \to K^0 p X$	& $>100$   & No \thplus seen \\
\noalign{\smallskip}\hline
\end{tabular}
\end{center}
\end{table}
\begin{table}
\caption{Publications with positive evidence for the \thplus~that 
have not yet been repeated with higher statistics.  As before, 
the statistical significance ($\sigma$'s) is that given in the 
published paper. (See the text for references.)}
\label{tab:2}       % Give a unique label
\begin{center}
\begin{tabular}{lccl}
\hline\noalign{\smallskip}
Group & Reaction & $\sigma$'s & Comment \\
\noalign{\smallskip}\hline\noalign{\smallskip}
ITEP  & $\nu A \to p K^0 X$	& 6-7 & 
NOMAD $\nu$ experiment also sees a small peak \\
ZEUS  & $e^+p \to p K^0 X$	& $\sim$4.5 &
new data with improved vertex detector is being analyzed \\
CLAS  & $\gamma p \to \pi^+K^-K^+n$	& 7-8 &
needs 5.5 GeV beam; might be rerun in 2008? \\
COSY  & $pp \to \Sigma^-K^0p$	& $\sim$5 &
new data with 5 times higher statistics is being analyzed \\
\noalign{\smallskip}\hline
\end{tabular}
\end{center}
\end{table}

\end{document}